\documentclass[acmsmall,screen,nonacm,review=false,timestamp=false]{acmart}
\AtBeginDocument{%
  \providecommand\BibTeX{{%
    \normalfont B\kern-0.5em{\scshape i\kern-0.25em b}\kern-0.8em\TeX}}}

\setcopyright{acmcopyright}
\copyrightyear{2018}
\acmYear{2018}
\acmDOI{XXXXXXX.XXXXXXX}

\acmConference[Conference acronym 'XX]{Make sure to enter the correct
  conference title from your rights confirmation emai}{June 03--05,
  2018}{Woodstock, NY}
\acmPrice{15.00}
\acmISBN{978-1-4503-XXXX-X/18/06}

\begin{document}

\title{Wikipedia in Wartime: Experiences of Wikipedians Maintaining Articles About the Russia-Ukraine War}

\author{Laura Kurek}
\email{lkurek@umich.edu}
\affiliation{%
  \institution{University of Michigan}
  \country{USA}
}
\author{Ceren Budak}
\email{cbudak@umich.edu}
\affiliation{%
  \institution{University of Michigan}
  \country{USA}
}
\author{Eric Gilbert}
\email{eegg@umich.edu}
\affiliation{%
  \institution{University of Michigan}
  \country{USA}
}

\begin{abstract}
How do Wikipedians maintain an accurate encyclopedia during an ongoing geopolitical conflict where state actors might seek to spread disinformation or conduct an information operation? In the context of the Russia-Ukraine War, this question becomes more pressing, given the Russian government’s extensive history of orchestrating information campaigns. We conducted an interview study with 13 expert Wikipedians involved in the Russo-Ukrainian War topic area on the English-language \textcolor{black}{edition of} Wikipedia. While our participants did not perceive there to be clear evidence of a state-backed information operation, they agreed that war-related articles experienced high levels of disruptive editing from both Russia-aligned and Ukraine-aligned accounts. \textcolor{black}{The English-language edition of} Wikipedia had existing policies and processes at its disposal \textcolor{black}{to counter such disruption}. State-backed or not, the disruptive activity created time-intensive maintenance work for our participants. Finally, participants considered \textcolor{black}{English-language} Wikipedia to be more resilient than social media in preventing the spread of false information online. We conclude by discussing sociotechnical implications for Wikipedia and social platforms.
\end{abstract}



\received{00 Month Year}
\received[revised]{00 Month Year}
\received[accepted]{00 Month Year}

\maketitle

\section{Introduction}
Wikipedia is one of the great successes of peer production and an essential information source on the internet~\cite{benkler_peer_2015, steinsson_rule_2023}. What happens, however, when Wikipedia encounters an ongoing, evolving geopolitical conflict? As Wikipedia is a community built on consensus, we seek to understand how Wikipedians adjudicate what belongs in a neutral encyclopedia article, when state actors on either side of a conflict might seek to promote their war-time messaging. 

In the context of the Russia-Ukraine War, this question becomes even more intriguing. The Russian government has a long history of spreading disinformation and conducting information operations, extending back into the Soviet era~\cite{rid_active_2020, lucas_winning_2016}. The word disinformation is etymologically related to the Russian word \textit{dezinformatsiya}, describing the intentional manipulation of the information environment to advance political goals~\cite{rid_active_2020}. Following the full-scale Russian invasion of Ukraine on February 24, 2022, an emerging body of scholarship is documenting how the Russian government has attempted to conduct information operations on various social media platforms, including Telegram and TikTok~\cite{aleksejeva_networks_2023, robinson_ukraine_2023, hanley_partial_2024}. The challenge of state-sponsored information campaigns is heightened in the case of the Russia-Ukraine War—making it a useful case study among the contentious topic areas on Wikipedia.

Theoretically, Wikipedia is an attractive target for an information operation, given its high visibility online and reputation as an authoritative information source. In 2022, the Wikipedia article titled ``Russian invasion of Ukraine'' was the second most viewed article on the English-language edition of the online encyclopedia, garnering over 50.6 million views~\cite{erhart_2022_2022}. The article remained in the top 30 most viewed Wikipedia articles in 2023, with over 12.7 million views~\cite{erhart_announcing_2023}. Wikipedia, however, is understudied in scholarly work on disinformation and information operations, with social media platforms often serving as the main object of inquiry. Furthermore, Wikipedia differs from social media in several important ways. First, social media sites prioritize social interaction and entertainment, while Wikipedia is engaged in knowledge production~\cite{mcdowell_it_2020}. Second, Wikipedia is perceived as an authoritative source of information, unlike social media~\cite{elmimouni_why_2022}. Third, Wikipedia is often served in the top results of an internet search.\footnote{\url{https://en.wikipedia.org/wiki/List\_of\_most-visited\_websites}} A Wikipedia article containing disinformation will arguably have a bigger impact than a social media post containing the same false information.

\subsection{Research questions}
In this work, we ask the following: 
\begin{itemize}
    \item[]  \textbf{RQ1}: \textcolor{black}{In the English-language edition of Wikipedia, did articles in the Russo-Ukrainian War topic area experience disruptive activity?} \vspace{3pt}
    \begin{itemize}
        \item \textbf{RQ1a}: \textcolor{black}{If so, did Wikipedians perceive there to be evidence of any state actor information operations targeting the Russo-Ukrainian War articles?} \vspace{3pt}
    \end{itemize} 
    \item[] \textbf{RQ2}: How well \textcolor{black}{did the English-language edition of} Wikipedia maintain articles related to the ongoing conflict of the Russo-Ukrainian War?
\end{itemize}  
To answer these questions, we conducted a semi-structured interview study with expert Wikipedia editors. We recruited 13 Wikipedia editors who maintained articles related to the Russo-Ukrainian War in the English-language edition of Wikipedia. For our sampling strategy, we first identified editors who had contributed extensively to the article ``Russian invasion of Ukraine,"\footnote{\url{https://en.wikipedia.org/wiki/Russian\_invasion\_of\_Ukraine}} and then identified additional Wikipedians based on these editors' recommendations, i.e. snowballing sampling. In the resulting sample, we heard from editors who had contributed to a wide range of the Russo-Ukrainian War articles as well as admins who assisted in the moderating this contentious topic area. 

\subsection{Summary of findings}
Across the 13 interviews, our participants \textcolor{black}{agreed that the Russo-Ukrainian War articles in the English-langauge edition of Wikipedia encountered high-levels of disruption from both Russia-aligned and Ukraine-aligned editors [\textbf{RQ1}]. Our participants described how the Russia-aligned and Ukraine-aligned editors employed similar disruptive tactics, but with differing informational goals.} Disruptive activity came from new, unregistered accounts as well as experienced accounts, with the latter having a greater impact in undermining information integrity on Wikipedia. Newer accounts were seen spamming talk pages---where article improvements and content disputes are discussed---with unconstructive criticism and personal attacks. Participant 13 drew a parallel between on the online conflict and the offline conflict: "People, just, they dig in deeper ... you get editors having trench warfare, just like it's actually happening." The more experienced accounts, meanwhile, exhibited a greater knowledge of Wikipedia's policies and processes. As such, some experienced editors engaged in wiki-lawyering---the misapplying of policies to one's benefit---to argue for articles to present their side in a more favorable light.

Our participants, however, did not perceive there to be conclusive evidence of a state-backed information operation targeting the Russo-Ukrainian War articles \textcolor{black}{[\textbf{RQ1a}]}. Participants did not consider there to be obvious signals of coordination, as encountered in other contested areas on Wikipedia. We explore potential reasons as to why this may be, given that social media platforms have been targeted by Russian state-aligned information campaigns. Regardless of whether a state actor was involved, \textcolor{black}{responding to the disruptive activity was time-intensive and tedious for our participants. Several participants expressed feeling burnt out or frustrated. As Participant 08 noted, "I’m kind of willing to spend some time to do that, but I’m not willing to get kicked in the teeth for it."}

To respond to this disruption, \textcolor{black}{the English-language edition of} Wikipedia had existing policies and processes at its disposal--honed in prior content disputes and contentious topics--\textcolor{black}{and editors employed some of the strictest protections \textbf{[RQ2]}.} For example, prior to the invasion, Wikipedia had already achieved consensus on disallowing most Russian state media for article citations: as Participant 07 put it, "we go by ‘how often this does this source tell outrageous lies?’" \textcolor{black}{The response also} involved applying extended confirmed (EC) page protections--which allow only editors with over 500 edits and a minimum of 30 days on Wikipedia to edit articles--to the entire Russo-Ukrainian War topic area: \textcolor{black}{such restrictions are seen in only four other topic areas in the English-language edition.} Participants agreed EC protections reduced disruptive editing, but did not fully eradicate it. State-backed or not, disruptive activity created a large mess.

Finally, participants described \textcolor{black}{the English-language edition of} Wikipedia as more adept than social media at impeding the spread of false information online. Participants considered this to be the case for the Russo-Ukrainian War, as well as other geopolitical conflicts. Participants noted that Wikipedia was an attractive target for an information campaign, just as social media is. Many thought that various aspects of Wikipedia---barriers to entry for editors and community-created polices---made the online encyclopedia resilient. Participant 12 explained, "Our processes come from the community and they’re operated by the community. At Twitter, Facebook, or wherever, you know, that’s all done largely top down." We conclude with implications other platforms may draw from Wikipedia's resilience. 

\section{Related Work}
Next, we review four branches of related research: 1) foundational literature on Wikipedia broadly; 2) work surveying contentious topics on Wikipedia; 3) \textcolor{black}{prior research on state-sponsored information operations (SSIOs), and in particular, the Russian government's use of information operations.}; and 4) recent work on the Russo-Ukrainian War topic area on Wikipedia.

\subsection{Wikipedia, the free encyclopedia}
Wikipedia is one of the most successful examples of internet peer-production---where individuals self-organize to produce goods or services without managerial directives or monetary compensation \cite{benkler_peer_2015, benkler_peer_2017, steinsson_rule_2023}. Over the past two decades, the Wikipedia community created a variety of policies and guidelines to support the building of an online encyclopedia. Forte \& Bruckman \cite{forte_scaling_2008} describe how achieving consensus is a key organizing goal on Wikipedia. To facilitate consensus building among editors, Wikipedians have crafted community norms, formal policies, and technological infrastructure~\cite{butler_dont_2008}. These community policies signal what the Wikipedia community considers important, both to external and internal stakeholders. In \textit{Should You Believe Wikipedia?}, Bruckman \cite{bruckman_should_2022} discusses how Wikipedia exemplifies the social construction of knowledge, where editors add content that is supported by the consensus of mainstream reliable sources, such as peer-reviewed research or reporting from reputable news organizations. 

\subsection{Contentious topics on Wikipedia}
Yet, consensus does not arise without conflict. In Wikipedia parlance, a \textit{contentious topic} is a topic that has experienced heightened levels of disruptive editing, and as such, admins are empowered to enact additional restrictions.\footnote{\url{https://en.wikipedia.org/wiki/Wikipedia:Contentious\_topics}} Edit wars, where editors revert each others’ edits back and forth, involve heated arguments instead of constructive discussion to improve an article \cite{sumi_edit_2011, yasseri_dynamics_2012, sepehri_rad_towards_2011}. Edit wars often indicate a contentious topic, where factions of editors disagree over what content should be presented on the article, and the ensuing edit wars can negatively affect article quality. Yasseri et al.~\cite{yasseri_most_2013} for example, found that across 10 language editions of Wikipedia, contentious topics prone to edit warring often involve ``religion, politics, and geographical places.''

Hickman et al.~\cite{hickman_understanding_2021} used both observational data and interviews with Wikipedia editors to understand how the contentious topic of the Kashmir region was being presented across Hindi, Urdu, and English-language Wikipedias. While they found differences in article coverage, organization, and editors’ approach to collaboration, they observed that across all three languages, editors strove to maintain a neutral point of view (NPOV)---a Wikipedia core content policy---and to prevent political agenda pushing. Kharazian, Starbird, \& Hill~\cite{kharazian_governance_2023} conducted a comparison of Serbo-Croatian Wikipedias to understand how governance capture by far-right editors occurred on the Croatian edition, but not the Serbian edition. Through the development of an “insular bureaucratic culture”, a small number of editors took over the governance structure of Croatian Wikipedia, dismantling neutrality-supporting policies and introducing instead far-right narratives and disinformation~\cite{kharazian_governance_2023}.

\subsection{\textcolor{black}{State-sponsored information operations}}

\textcolor{black}{An information operation describes an organized attempt to manipulate the information environment towards a strategic goal \cite{arif_acting_2018, starbird_disinformation_2019, polychronis_working_2023}.  In the 21st century, governments and political actors have leveraged the internet as another medium through which to conduct such operations \cite{bradshaw_global_2019}. Information operations can involve the propagation of true, false, speculative, or misleading content -- making it a broader term than disinformation, which describes the intentional spreading of falsehoods \cite{starbird_disinformation_2019, beers_demographics_2022}. CSCW scholarship has emphasized the collaborative nature of information operations online: while such operations are typically orchestrated by governments, the participation of human crowds online, or "unwitting agents" \cite{bittman_kgb_1985}, can be central to the operation's success \cite{starbird_disinformation_2019, polychronis_working_2023}.}

\textcolor{black}{The increasing prevalence of information operations online has led to a large body of scholarship on their detection. This detection work has predominantly focused on the social media platform X (formerly known as Twitter) and has explored various approaches to identify coordinated behavior online --- an indicator of a information campaign \cite{davis_botornot_2016, cresci_paradigm-shift_2017, shao_anatomy_2018, pacheco_uncovering_2021, ferrara_disinformation_2017}. Detection work related to  Wikipedia is often focused on sockpuppet accounts --- where an editor user misuses multiple accounts, often to deceive other editors or evade bans \cite{kumar_army_2017, sakib_automated_2022}.}

\textcolor{black}{Our work offers a departure from much of the online information operations literature to date. Instead of attempting to quantitatively detect an information operation, we conduct qualitative interviews with platform moderators and users to understand their mental model of what might constitute an information operation on their platform --- in this case the crowd-sourced encyclopedia Wikipedia.}

\subsubsection{Russian state-sponsored information operations and disinformation} 
The Russian government’s use of information operations and disinformation for domestic, near-abroad, and international audiences is well-studied. Russia’s focus on information as a strategic instrument can be traced to the `active measures' of the Soviet era, where intelligence services attempted to manipulate the information environment to influence political outcomes~\cite{rid_active_2020}. In the 21st century, Russia has continued the Soviet legacy of information manipulation using the scale and anonymity of the internet, often targeting post-Soviet and post-communist countries, such as Estonia, Latvia, Poland, and Ukraine~\cite{lucas_winning_2016}. Following the 2013-2014 Euromaidan protests and the ensuing invasion of Crimea, the Russian government created low-quality news sites and social media accounts to delegitimize Ukraine's interest in partnering closer with the European Union~\cite{lucas_winning_2016, jankowicz_how_2020}.

Beyond Central and Eastern Europe, a Russian information operation targeted the 2016 U.S. presidential election, with the creation of fake accounts on Facebook and Twitter to engage in political debates and stoke division among U.S. voters~\cite{arif_acting_2018}. Further investigation revealed a Russian state-supported organization known as the Internet Research Agency (IRA) was behind the 2016 U.S. election information operation~\cite{carroll_st_2017, dossier_cyber_2023}. A large body of research has sought to understand the activities of the IRA `trolls' during the 2016 U.S. election, studying the thematic content of the imposter social media accounts and the interactions that these inauthentic posts garnered from actual people~\cite{badawy_analyzing_2018, diresta_tactics_2019}.

Following the 2022 Russian invasion of Ukraine, research is emerging to understand how Russia has attempted to manipulate the information environment in yet another geopolitical conflict. \textcolor{black}{Current scholarship has analyzed Russian state media messaging on various social media platforms: Twitter \cite{park_challenges_2022, pierri_propaganda_2023}, Facebook \cite{pierri_propaganda_2023}, Reddit \cite{hanley_happenstance_2023}, Telegram \cite{aleksejeva_networks_2023, hanley_partial_2024}, and VKontakte \cite{park_challenges_2022}. Studies have sought to describe the messaging strategies of the Russian government \cite{park_challenges_2022}, as well as track how these narratives move across the internet \cite{hanley_partial_2024}. Even with social media platforms placing restrictions on Russian state media accounts following the 2022 invasion, one study found that Russian state-aligned messaging continued to spread on Facebook and Twitter \cite{pierri_propaganda_2023}.}

Given that research on Russian disinformation and information operations around the war in Ukraine has focused primarily on social media platforms, we seek to extend the literature beyond social media to include knowledge-production platforms like Wikipedia.

\subsection{The Russo-Ukrainian War on Wikipedia}
As a recent geopolitical conflict, there is limited research on the Russo-Ukrainian War on Wikipedia. Roberts \& Xiong-Gum~\cite{roberts_wikipedia_2022} conducted a content analysis of the edit history of the article "Russian invasion of Ukraine" from February 24, 2022 to March 2, 2022. The authors described how editors acted as vandal fighters by reverting disruptive edits and argued that the editors' actions exhibit connective intelligence, whereby “editors connect with others toward a common goal”~\cite{roberts_wikipedia_2022}. The work also notes that the article’s infobox as a site of conflict, where editors disagreed over which countries to include as belligerents---given the military equipment supplied to Ukraine by various countries.

Dammak \& Lemmerich~\cite{dammak_effects_2023} conducted an observational study of articles related to the Russo-Ukrainian war on the Russian, Ukrainian, and English-language editions. In the initial week following the invasion, they found an increased revert rate across all language editions---an indication of elevated conflict and disagreement on these pages. The elevated levels of reverting, however, returned quickly to normal rates within two weeks, which the authors surmise was a result of additional editing restrictions and increasing consensus on the articles. 

Prior to the 2022 invasion, Kozyr \& Dubina~\cite{kozyr_wikipedia_2017} compared the Ukrainian and Russian language editions' coverage of the war in Donbas, which has been ongoing since Russia’s invasion of Crimea in 2014. They describe an “informational struggle” evident between the two Wikipedias, with article titles differing in their description of the conflict, e.g. "War in Eastern Ukraine" in the Ukrainian edition versus "Armed conflict in Eastern Ukraine" in the Russian edition.

Our paper contributes to the body of existing Wikipedia research on contested topics by conducting the first interview study of Wikipedians involved in the Russo-Ukrainian War topic area. In contrast to previous work in this space, we also investigate whether these articles were targeted by a state-sponsored information operation.

\section{Method}
We use a qualitative research design. We conducted 13 interviews with expert Wikipedians who edit articles about the Russia-Ukraine war. The study was approved by our university's Institutional Review Board (IRB). An interview study has several advantages in this context. First, as Wikipedia runs on a complex sociotechnical ecosystem of policies and processes created by the community, we wanted to hear from editors firsthand about how contentious topics are handled and which policies they employ. Second, to understand where the problem areas were, we considered it more effective to ask editors directly, rather than attempting to reverse-engineer the conflicts via edit history logs. Editors would be able to tell us the story behind the edit wars, where else to look, and who else to talk to. Third, interviewing editors would provide a sense of how the Russo-Ukrainian war topic area compares to other contentious topic areas.

\subsection{Recruitment}
We used a purposeful and snowball sampling strategy to recruit expert Wikipedia editors closely involved with articles related to the war in Ukraine. We selected the article "Russian invasion of Ukraine" (RIU) as our seed article, as it is a `parent' article which links to other `child' articles on the war.\footnote{\url{https://en.wikipedia.org/wiki/Russian\_invasion\_of\_Ukraine}} We utilized the website XTools to collect the top 40 editors by edit count on the RIU article talk page---where content disputes and article improvements are discussed.\footnote{\url{https://www.mediawiki.org/wiki/XTools}} XTools provides statistical summaries of article history and editor activity on Wikipedia. With the list of 40 editors, we examined their editing history using both raw edit logs from Wikipedia and user contribution summaries from XTools.

Following our analysis of editor contributions, we narrowed our recruitment list to 22 editors who met the following criteria: (1) consistent engagement with the RIU article or other war-related articles (contributing edits month over month, following the influx of editing in the first weeks of invasion), (2) constructive discussion on the RIU talk page (as assessed by the first author), and (3) indication of willingness to be contacted. We considered an editor amenable to contact if their User page had an "Email this user" link and/or included text inviting contact from other editors. Of the 22 Wikipedia editors that met the above three criteria, 10 agreed to be interviewed. We asked these 10 editors for recommendations on who else to talk to, i.e., snowball sampling, resulting in three additional interviews. 

\subsection{Participants}
With this combination of purposeful and snowball sampling, we conducted 13 interviews across five time zones. In our purposeful sampling, our coverage is high: 10 of the 22 criteria-passing editors agreed to be interviewed. For snowball sampling, when we asked participants who else to recruit for the study, our participants frequently mentioned each other (i.e., already interviewed Wikipedians). Among the recruited editors, most had contributed to a variety of Russo-Ukrainian War articles, beyond the main RIU article. In the resulting sample, we were able to hear from editors involved in a wide range of  war-related articles as well as admins who moderate the Russia-Ukraine topic area. Table 1 lists the participants, their roles, and Wikipedia experience as summarized by number of edits and years on the platform. As Wikipedians do not supply demographic information to the wider internet, we did not collect demographic information to ensure the anonymity of our participants. 

\begin{table}[]
\begin{tabular}{lllr}
\textbf{Participant ID} & \textbf{Role}      & \textbf{No. of edits} & \multicolumn{1}{l}{\textbf{Editor since}} \\ \midrule
P01                     & extended confirmed & 50K-60K               & 2016                                      \\
P02                     & extended confirmed & 10K-20K               & 2013                                      \\
P03                     & extended confirmed & 20K-30K               & 2021                                      \\
P04                     & extended confirmed & 10K-20K               & 2011                                      \\
P05                     & extended confirmed & 10K-20K               & 2014                                      \\
P06                     & extended confirmed & 60K-70K               & 2010                                      \\
P07                     & extended confirmed & 60K-70K               & 2007                                      \\
P08                     & extended confirmed & 90K-100K              & 2006                                      \\
P09                     & extended confirmed & 10K-20K               & 2019                                      \\
P10                     & admin              & 200K-300K             & 2011                                      \\
P11                     & admin              & 20K-30K               & 2009                                      \\
P12                     & admin              & 200K-300K             & 2005                                      \\
P13                     & extended confirmed & 10K-20K               & 2021                                      \\ \hline
\end{tabular}
\caption{List of interview participants}
\label{Table 1}
\end{table}

\subsection{Interview procedure}
All interviews were conducted by the lead author remotely over video conferencing software between October 2023 and January 2024. The length of interviews ranged from 57 minutes to 2 hours and 26 minutes, with an average of 1 hour and 33 minutes.\footnote{Interviewees often agreed to continue talking beyond the allocated time of 60 minutes.} Participants were compensated with \$30 for participating in the interview. All interviews were recorded and transcribed using transcription software. The semi-structured interview questions were designed to draw out editors' experiences maintaining articles related to the war in Ukraine. Editors were asked (1) how they got involved in the Russo-Ukrainian War topic area, (2) what issues did the war-related articles face, and (3) how well did Wikipedia policies and processes support the resolution of these issues. See Appendix A for the full set of interview questions. After the first three interviews, the authors met to review the initial findings and revisit the interview protocol. We had begun to notice diverging opinions among editors as to how Wikipedia was responding to the contested war-related articles. We updated the protocol to include questions to further understand (4) what each editor's focus was on Wikipedia (e.g., Eastern Europe versus Military History) and (5) how editors' judged the intent of disruptive accounts.

 \subsection{Data analysis}
The first author reviewed each auto-generated transcript and made corrections as needed via the transcription service Otter.AI. The interview transcripts were analyzed using three rounds of qualitative coding inspired by thematic analysis \textcolor{black}{-- an inductive, bottom-up approach which considers multiple observations across the collected data to derive codes and higher-level themes}~\cite{saldana_coding_2013}. Using MAXQDA software, the first author conducted open coding and In Vivo coding for the first round. In the second round, focused coding was employed to identify emerging concepts from the initial codes. Examples of second-round codes included "main versus periphery articles," "articles stabilize over time," and "assume good faith culture." First and second round coding proceeded in an iterative fashion, shifting between data collection and qualitative analysis as common concepts coalesced from participant responses. The first author engaged in memo writing to keep track of key themes and patterns. 

\textcolor{black}{The iterative rounds of open coding resulted in hundreds of codes. The rounds of focused coded synthesized the initial codes into roughly 100 concepts across the 13 interviews.} In the third round of analysis, the first author constructed an affinity map of the focused codes using MAXQDA. \textcolor{black}{The affinity map distilled the focused codes into higher-level themes, such as "Activities of Russia-aligned editors," "Activities of Ukraine-aligned editors," "Disruption on Wikipedia," "Coordinated activity on Wikipedia", "Reliable sources policy," and "Page protections policy."} All authors then met to synthesize the themes most central to the research questions. \textcolor{black}{Finally, the authors structured the central themes into three high-level findings, each with several sub-findings nested beneath. Each high-level finding corresponds to a research question: RQ1, RQ1a, and RQ2.}

\subsection{Methodological limitations}
Our sampling strategy and qualitative approach also have limitations. As a purposeful/snowball design, we can make limited claims about the representativeness of the reports given by our interviewees. Second, our qualitative design does not permit large-scale inferences about how Wikipedia deployed its sociotechnical architecture in a time of conflict. 

\section{Findings}

\subsection{\textcolor{black}{In the English-language edition of Wikipedia, Russo-Ukrainian War articles encountered disruption from both Russia-aligned and Ukraine-aligned editors.}}
\textcolor{black}{In answer to RQ1,} the 13 Wikipedia editors we interviewed described how the RIU article and other articles related to the Russo-Ukrainian war experienced high levels of disruptive editing \textcolor{black}{from both Russia-aligned and Ukraine-aligned editors}. The participants, however, did not perceive there to be evidence of a state-backed information operation from either Russia or Ukraine. Wikipedia has established norms around what is considered constructive editing behavior, which are encapsulated in policies, guidelines, and explanatory essays. The behavioral guideline of \textit{disruptive editing} considers disruption to include: persistent pushing of a point of view, failure to cite reliable sources, failure to build consensus with other editors, and ignoring community feedback.\footnote{\url{https://en.wikipedia.org/wiki/Wikipedia:Disruptive\_editing}} In our interviews, participants described encountering users who engaged in disruptive editing to push both Russia-aligned and Ukraine-aligned points of view onto articles related to the Russo-Ukrainian War. 

Participants described how disruptive editing occurred shorty after the RIU article's creation. Disruptive editors vandalized the infobox template parameters, changing the conflict title from "2022 Invasion of Ukraine" to "2022 Liberation of Ukraine." Other disruptive edits targeted the infobox's conflict status: one editor changed the status to "Resolved," while another editor added extraneous text aligned with Russian state messaging that described the conflict as "a military occupation with the goal of demilitarization and denazification." Within 12 hours, the RIU article was protected so that only editors with over 500 edits and over 30 days on Wikipedia could edit the article, known as extended confirmed protection (ECP).\footnote{\url{https://en.wikipedia.org/wiki/Wikipedia:Protection\_policy}} Page protections will be discussed in a later section, as a central tool used by Wikipedia editors to maintain contested articles. 

Following the article page protections, participants recalled how disruptive editing continued on the talk pages, which are intended as a space to resolve disagreements over article content. Both Russia-aligned and Ukraine-aligned editors used talk pages to push their point of view, rather than constructively debate article improvements. Participant 05 noted, "... you get a lot of inexperienced editors with opinions that are very pro-Russian or very pro-Ukrainian. Not neutral. And make drive by statements". Participant 07 recalled:
\begin{quote}
    \textit{"It became a battlefield. As obviously, pro-Russian editors were trying to say how brilliantly the Russians were doing. Pro-Ukrainian editors were saying how badly the Russians are doing."}
\end{quote} This state-aligned point of view pushing on both sides was in violation of neutral point of view---one of Wikipedia's core content policies---which requires that articles "must not \textit{take} sides, but should \textit{explain} the sides".\footnote{\url{https://en.wikipedia.org/wiki/Wikipedia:Neutral\_point\_of\_view}} 
The scale of disruptive editing can also be seen in the length of the talk page archives. Prior research has shown that the size of talk page archives can indicate conflict on Wikipedia~\cite{sumi_edit_2011, yasseri_dynamics_2012}. Once a disagreement has been resolved, editors are expected to archive the discussion, allowing the talk page to remain navigable and present only current debates. The longer the talk page archives, the more disagreements that have occurred. Participant 07 described how the RIU article accumulated over five talk page archives only one week into the conflict: "We have basically pretty much ... almost a live update." Participant 04 recalled archiving a large number of talk page discussions "because of the sheer volume of people coming in." 

\subsubsection{\textbf{Russia-aligned editing activity differed from Ukraine-aligned editing activity in terms of goals, but not necessarily in terms of tactics.}} 
Participants reported that both Russia-aligned and Ukraine-aligned accounts engaged in disruptive editing on articles related to the war. We find that both types of accounts tended to employ similar disruptive tactics, while the respective informational goals of their activity differed. \textcolor{black}{We discuss two tactics in particular -- wiki-lawyering and creation of low-quality articles} -- and compare Russia-aligned goals to Ukraine-aligned goals.

\textbf{Tactic: Wiki-lawyering} The term wiki-lawyering is used by the Wikipedia community to refer to the misapplying of policies to one's benefit.\footnote{\url{https://en.wikipedia.org/wiki/Wikipedia:Wikilawyering}} Our participants explained that a clear example of wiki-lawyering would be to use a certain interpretation of a policy in one dispute and then use a different interpretation of the same policy in another dispute. The phenomenon of wiki-lawyering has only been given passing attention in existing work~\cite{ford_anyone_2017, de_kock_i_2021}, so we discuss it in greater detail here.

\textit{Example of \textcolor{black}{Russia-aligned wiki-lawyering:}} From the examples provided by our participants, we found that Russia-aligned accounts appear to use wiki-lawyering to delegitimize Ukraine as a nation state. Participant 06 described how Russia-aligned editors would try to insert the full title for Russian president Vladimir Putin, but not for Ukrainian president Volodymyr Zelenskyy, instead updating the RIU article to read "Zelenskyy." Participant 06 recounted: 
\begin{quote}
    "\textit{I spent many days trying to put an equivalence on the [article]. I gave up eventually. Again, it was wiki-lawyering ... `Once you've used the word president once, you don't use it a second time. There's no need.' It's yeah ... a delegitimization.}" 
\end{quote} Other editors recalled that Russia-aligned editors used wiki-lawyering on the Azov Brigade article to describe the Ukrainian volunteer military unit as definitively neo-Nazi affiliated, while attempting to minimize the neo-Nazi affiliations of the Wagner Group, the Russian state-affiliated private military company. Participant 07 remembered, "There was a different standard being applied to Wagner Battalion as opposed to the Azov Battalion ... there was a greater willingness to say that the Azov Battalion were Nazi and not the Wagner Battalion." 

Participant 08 described wiki-lawyering in the article titled "Sexual violence in the Russian invasion of Ukraine." Several Russia-aligned editors argued repeatedly for the article to caveat negative reports about the Russian army, while advocating to include one report of sexual violence committed by the Ukrainian army. The Russia-aligned editors raised these issues several times in the article's talk page, citing Wikipedia policies such as neutral point of view and verifiability in attempt to support their case.\footnote{\url{https://en.wikipedia.org/wiki/Wikipedia:Neutral\_point\_of\_view}}\footnote{\url{https://en.wikipedia.org/wiki/Wikipedia:Verifiability}} Participant 08 recalled, "I don't see how anybody can in all good faith, you know, wiki-lawyer an article into saying that."  

\textit{Example of Ukraine-aligned \textcolor{black}{wiki-lawyering:}} Multiple participants mentioned the article "Battle of Bakhmut" as a site of conflict between Ukraine-aligned and Russia-aligned editors, especially after Russia's capture of the administrative boundaries of the city in May 2023.\footnote{\url{https://en.wikipedia.org/wiki/Battle\_of\_Bakhmut}} In the talk page, editors debated whether the status of the battle should remain as "ongoing". Several Ukraine-aligned editors argued that reliable sources did not describe the battle as an outright Russian victory and that fighting continued in the city's outskirts. In response to the battle's status remaining as "ongoing", the talk page received an influx of criticism from Russia-aligned accounts. Participant 01 recalled, "It attracted a lot of pro-Russians on the talk page…saying that you just cannot accept Russian victory, that you are pushing Western propaganda." Several participants considered part of the issue was that Western reliable sources were at times sympathetic to Ukraine. Participant 09 commented, "There is a persistent pro-Ukrainian bias in a lot of places where, with like the counter offensive ... it'll fail, and Wikipedia cannot say that, because sources don't say it. They don't want to say it."  

Another example of Ukraine-aligned wiki-lawyering occurred in the article for Donetsk, a city in eastern Ukraine, which has been occupied by Russian-backed militants since 2014. In September 2022, Russian president Vladimir Putin announced the annexation of Donetsk and the surrounding region---an act decried as illegal by Ukraine and other nations~\cite{trevelyan_as_2022}. Despite this political turmoil, Participant 10 noted that in the Donetsk article, the infobox listed the country as Ukraine without mentioning Russia. In the first paragraph, Russia's 2014-present occupation is mentioned, but not the 2022 annexation, which is mentioned only later in the article. In the article's talk page, one editor requested that the infobox display the de facto country as Russia and the de jure country as Ukraine---a compromise seen in the article for the city of Sevastopol in Crimea. In response, one Ukraine-aligned editor argued (1) that there were no reliable sources to support this edit and (2) that the Wikipedia Manual of Style advises that an article's first sentence should not be overloaded with information. Using a content policy and a manual of style guideline, the Ukraine-aligned editor prevented the mentioning of the 2022 Russian annexation.\footnote{\url{https://en.wikipedia.org/wiki/Wikipedia:Manual\_of\_Style/Lead\_section\#First\_sentence}} Participant 10 recalled:
\begin{quote}
    "\textit{They claim ... there are no reliable sources saying [Donetsk] is annexed because Russian sources on that cannot be reliable. Which to some extent makes sense. But I mean, at the end of the day, right, we are there to provide information. We are not there to fight on technicalities. And that sometimes people forget.}"
\end{quote}

\textbf{Tactic: Creating articles that do not meet Wikipedia's standards} Wikipedia uses the idea of notability---topics with sizable coverage by reliable sources---to determine what warrants an article on the online encyclopedia.\footnote{\url{https://en.wikipedia.org/wiki/Wikipedia:Notability}} Additionally, Wikipedia does not intend to act as a newspaper nor a public relations agency.\footnote{\url{https://en.wikipedia.org/wiki/Wikipedia:What\_Wikipedia\_is\_not}} Participants recounted how Russia-aligned and Ukraine-aligned accounts created articles that did not meet Wikipedia's standards. 

\textit{\textcolor{black}{Example of Russia-aligned low-quality articles}} Participant 09 and Participant 13 described a Russia-aligned editor who engaged in a pattern of creating low-quality articles for small battles and skirmishes in the Russo-Ukrainian War. The articles would contain extraneous details, employ the Russian spelling of geographical places, and often mis-cite sources. When others sought to delete or merge these articles, the Russia-aligned editor would undo their actions. Participant 09 noted that responding to these low-quality articles, especially checking poor citations, was a time-consuming task. Another Russia-aligned editor created the article "Alley of Angels" describing a war monument in Russia-occupied Donetsk. The article was only three sentences long, provided no citations, and claimed that the Ukrainian military had killed over 500 children. After being approved for deletion in October 2022, the page was soon recreated in November 2022.\footnote{\url{https://en.wikipedia.org/wiki/Talk:Alley\_of\_Angels}} At a UN Security Council briefing in December 2022, Russian Ambassador Vasily Nebenzya used the article's deletion to claim that the West is concealing the truth of the war in Ukraine~\cite{shcherba_ambassador_2022}. Participant 09 recalled, "After he made that statement, there was an influx of like IP accounts to the talk page of the deleted article, saying, `Why is there no article? Nothing will ever be good enough for you.'" This article was ultimately removed in February 2023.\footnote{\url{https://en.wikipedia.org/w/index.php?title=Alley\_of\_Angels\&redirect=no}}

\textit{\textcolor{black}{Example of Ukraine-aligned low-quality articles}}  Ukraine-aligned editors, meanwhile, attempted to document each moment of the war. In the RIU article, Ukraine-aligned editors would provide daily updates on the unfolding invasion. These updates grew so numerous that editors decided to separate out these daily updates into timeline articles, with ultimately six timeline articles created by December 2023. Participant 07 remembered:
\begin{quote}
    "\textit{So every single time a Russian tank was knocked out, it'd be put in the article ... Every time the Russians took a village, it'd be put in the article.}"
\end{quote} Several participants considered such content to not meet Wikipedia's standard of notability, given its similarity to live updates from a news outlet and use of emotional tone. Participant 06 noted that with the iterative cycles of editing, the articles would attain a more sober tone: "In the fullness of time, all of that trivia ... all of that emotion gets whittled out."

\subsection{\textcolor{black}{Wikipedia editors did not consider the Russo-Ukrainian War disruption to be part of a state actor information operation.}}

\textcolor{black}{In answer to RQ1a, the 13 Wikipedia editors we interviewed did not perceive there to be evidence of a state-backed information operation from either Russia or Ukraine.} Participants overall were hesitant to label the disruptive editors as part of a state-backed information campaign. This hesitancy extended to the Russia-aligned accounts, despite reports of Russian state media narratives circulating on social media as well as Russia's track record of waging information campaigns against Ukraine since the 2014 invasion of Crimea~\cite{pierri_propaganda_2023, lucas_winning_2016}. Participants did not perceive the Russia-aligned accounts to be coordinated \textit{enough}, as seen in other contested topic areas on Wikipedia, e.g., Israel-Palestine, Nagorno-Karabakh, or India-Pakistan. Participant 04 described the editing activity on the RIU article as such: \begin{quote}
    "\textit{It didn't really feel coordinated ... If they were part of some kind of government, they didn't seem to be well coordinated or run ... like, if we're talking about an organized disinformation campaign, I would say at best it would be a disorganized information campaign.}"
\end{quote}

\subsubsection{\textbf{Disruptive activity in Russo-Ukrainian War articles did not resemble other activity considered to be likely state-backed coordination on Wikipedia.}}

Participant 06 recalled working on articles related to the Nagorno-Karabakh conflict between Armenia and Azerbaijan and described crude coordination from the Azeri side. The most clear signals of coordination included recruiting potential editors on Facebook, copy-pasting wording from government press releases, and uploading high-quality photos of the President of Azerbaijan which presumably could only have been taken by an official press pool. Additional signals of coordination involved an “unlimited supply of editors on the Azeri side” who did not mind being banned or blocked, “a whole scale Azerification of names, places, everything” on even the most obscure of articles, and “cloying praising” of the Azerbaijani government. In comparing Azeri-aligned activity with Russia-aligned activity, Participant 06 noted the Russia-aligned editors did not engage in "hero worship" of the Russian government. Additionally, while Russia- and Ukraine-aligned editors engaged in edit warring over the spelling of geographic place names as Azeri- and Armenian-aligned editors did, Participant 06 considered only the Azeri-aligned disruptive editing to be “relatively transparent, bad coordination of state actors.” Participant 06 added: \begin{quote}
    "\textit{I'm 100\% convinced that there were state actors involved in the Artsakh [Nagorno-Karabakh] campaign. I'm not so convinced on Russia.}"
\end{quote}

Participant 10 suspected Russia-aligned editors active in 2014-2016 of being paid, given their editing exclusively during working hours, poor English skills, and appearance of having a “clear agenda” from which they did not deviate. Paid editing on Wikipedia must be disclosed. Otherwise, the offending accounts can be indefinitely blocked. Wikipedia maintains a list of paid editing companies known for violating this Wikipedia policy.\footnote{\url{https://en.wikipedia.org/wiki/Wikipedia:List\_of\_paid\_editing\_companies}} The 2014-2016 Russia-aligned editors would consistently go from article to article to make an identical edit, such as changing Donetsk from a Ukrainian city to a Russian city. When these editors were confronted about their disruptive behavior, they would not engage in talk page discussions to defend their edits, nor would they protest if topic banned or blocked. Another account would simply appear and begin making similar edits. Participant 10 contrasted this with Ukraine-aligned editors who would protest an account block and attempt to repeal it: 

\begin{quote}
    \textit{"But yeah, [Ukraine-aligned editors in 2014-2016] were very annoying. [Russia-aligned editors in 2014-2016] were not annoying. I mean, they were just doing their job. You block, you revert, they don't come back ... You see that one side is editing because they ... feel that they should edit. Another side is editing because they're just paid to edit."}
\end{quote}By contrast, Participant 10 did not consider the disruptive editing occurring on the Russo-Ukrainian war articles following the 2022 invasion to be from paid editors on either the Russia-aligned or Ukraine-aligned side.

\subsubsection{\textbf{\textcolor{black}{Wikipedia} editors suspect that many of the Russia-aligned accounts could be individuals who consume state media as opposed to state-paid trolls.}}
Many of the participants that we interviewed acknowledged that Wikipedia is an attractive target for information manipulation by governments, companies, and other actors. Often cited was Wikipedia's high placement in Google search results. Participant 08 noted: 
\begin{quote}
    "\textit{If you're looking at it from the point of view of a search engine optimization, you really can't do better than Wikipedia. Because if you want to control a narrative, Wikipedia articles are usually the first or the second, you know, thing returned.}" 
\end{quote}

While several participants surmised that the Russian or Ukrainian governments could have been involved in the disruption surrounding the war-related articles, none said that they had any direct evidence of such involvement. Participants, instead, tended to assess the disruptive Russia-aligned editors as more likely to be private individuals with nationalistic, pro-Russia views rather than paid, state-backed trolls. Participants considered discerning the intent of a disruptive editor to be difficult, often surmising that Russia-aligned editors might be individuals who consume primarily Russian state media. 
\begin{itemize}
    \item P01: "\textit{I am inclined to think that this is just individual, independent people who have a pro-Russian position and who genuinely doubt certain things that are said to happen on the war, which is very clear proof of the effect of Russian disinformation.}"
    \item P02: "\textit{I would say it's very hard to distinguish between actors who are probably being financed by the Russian state to insert misinformation, and people who are maybe just watching Russia Today and being a bit brainwashed}."
    \item P06: "\textit{I think it may have been just, you know, ultra nationalism, but private, ultra nationalism}."
    \item P07: "\textit{Whereas a lot of Russian users, obviously because they've been fed the state media, do definitely see themselves as being the victims of Western propaganda.}"
    \item P08: "\textit{The people that are arguing for Russia ... I think some of it is simply that they consume different media. You know, they don't believe the Western media.}"
    \item P09: "\textit{I wouldn't say that they're like a Russian bot, that like they are working for the Internet Research Agency. But I would say this is bad faith. I would say that they know they're not in line with policy…But to me this just seems like a rage that [Wikipedia] is supposedly biased against Russia and knowingly breaking the rules of the website to try to remedy that.}"
\end{itemize}

Wikipedia's behavioral guideline "Assume good faith" encourages editors to assume others are acting in good faith to build an encyclopedia, even if their editing is disruptive.\footnote{\url{https://en.wikipedia.org/wiki/Wikipedia:Assume\_good\_faith}} Participants frequently referred to this norm of assuming good faith when describing the disruptive editing they had encountered. Participant 04 noted that on talk pages of Russo-Ukrainian War articles, disruptive editors often criticized the article without proposing any improvements. Participant 04 considered that these editors were either unaware of Wikipedia's policies or that they did not "really care how the encyclopedia works." Other participants similarly described disruptive editors as new and inexperienced. 

For disruptive editors who continue disrupting even after being warned, the Wikipedia community has developed a lexicon to describe various problematic behaviors. One of these terms is "point of view pushing" or "POV pushing," referring to neutral point of view---one of Wikipedia's core content policies.\footnote{\url{https://en.wikipedia.org/wiki/Wikipedia:NPOV\_dispute\#POV\_pushing}} Another term is "single purpose account," which describes an account whose editing is concentrated on a small set of articles and appears to promote an agenda.\footnote{\url{https://en.wikipedia.org/wiki/Wikipedia:Single-purpose\_account}} There are also terms to describe coordinated editing across multiple accounts: sockpuppet, where one user attempts to edit under various names, and meatpuppet, where multiple users are recruited for disruptive activities.\footnote{\url{https://en.wikipedia.org/wiki/Wikipedia:Sockpuppetry}} 

For the editors we interviewed, ultimately, it did not matter whether a disruptive editor was suspected to be paid by a state actor or not, as the disciplinary actions taken were the same. With terms like single purpose account and sockpuppet, Wikipedians appear to be focused on drawing out behavior that obstructs the building of the encyclopedia, rather than determining whether the problematic behavior is coming from opinionated individuals or paid state actors. Participant 07 noted, "So whether it be state organized or privately organized. It's the same thing. It's meatpuppetry." Participant 09 expressed a similar sentiment:
\begin{quote}
    "\textit{[It] doesn't really matter if they're a state actor or not. Because the results are basically the same as if they're a lone actor.}" 
\end{quote}

\subsubsection{\textbf{\textcolor{black}{State-backed or not, disruptive activity is time-consuming for Wikipedia editors to address.}}} \textcolor{black}{Regardless of whether a state actor was involved or not, disruptive activity---whether from new IP users or experienced wiki-lawyering editors---created additional work for Wikipedia editors seeking to maintain a reliable encyclopedia during an ongoing war. On the talk pages, unconstructive comments and personal attacks needed to be archived, and offending users required disciplinary action---typically a warning at first and then later a ban or block of some kind. Participant 07 recalled responding to disruptive activity on the RIU article's talk page:} 
\begin{quote}
    "\textit{We were just being bombarded with essentially unactionable requests, which was just wasting our time, because you have to read them. Because if you don't read them, you can't know they are actionable. And that takes time. It's basically time wasting.}"
\end{quote} \textcolor{black}{For low-quality articles, citations needed to be double checked or replaced entirely. Participant 09 recounted working on these low-quality articles:} 
\begin{quote}
    "\textit{It was making basically a complete nightmare for people who wanted to like edit it. Because to fix things, I would have to go through to every claim, check the source. And oftentimes it just wouldn't say what [the Russia-aligned editor] was saying.}"
\end{quote}

\textcolor{black}{When recalling their efforts at countering disruption, participants often expressed feelings of burnout and frustration. Participant 06 commented, "Sometimes you have the energy to combat them. Sometimes you don't." Participant 08 added, "I'm kind of willing to spend some time to do that, but I'm not willing to get kicked in the teeth for it." Participant 10 said, "I have only so much time, I can spend it elsewhere." On a more positive note, Participant 11 explained that extended confirmed page protections can assuage editor burn out: "The editors sort of who are here in good faith are more likely to stay and are less likely to just burn out knowing that there's just this general level of protection."}

\textcolor{black}{Several participants also noted that the experienced Russia-aligned editors would attempt to retaliate against editors who called out their policy-violating conduct. Participants recalled how they were reported to administrator noticeboards, being called an `agenda-pusher' by the very agenda pushers they were trying to stop. Participant 03 described the Russia-aligned EC editors: "They're very clever. They know what they're doing. They're familiar with Wikipedia policy. They know how to, if they get in trouble, play victim."}

\subsection{\textcolor{black}{English-language Wikipedia had policies in place to protect against disruptive activity, and editors employed some of the strictest protections.}} \textcolor{black}{In answer to RQ2,} throughout our interviews, participants' responses were replete with references to myriad Wikipedia policies, guidelines, and informational essays. Our participants—unpaid, volunteer editors—appeared to understand the policies front-to-back and how to apply them. Some of the most intensive editing restrictions were employed to maintain information integrity on the Russo-Ukrainian War articles. Our participants, however, noted that peripheral articles still suffered from disruptive activity, given editors' limited attention and time. Our participants were split in their perception of how Wikipedia responded to the war-related disruption overall.

\subsubsection{\textbf{\textcolor{black}{Wikipedia editors relied on page protections, talk page moderation, and reliable sources to maintain Russo-Ukrainian War articles.}}} \textcolor{black}{Among the policies available to Wikipedia editors to counter disruptive activity, some of the most intensive editing restrictions were employed. These restrictions involved both limiting who could edit articles related to the Russo-Ukrainian War and moderating talk page discussions.}

\textit{Limiting who can edit articles:} In response to disruptive editing, articles related to the Russo-Ukrainian War received extended confirmed (EC) page protection as well as a contentious topic designation. Under EC protections, only EC editors (over 500 edits and over 30 days on Wikipedia) can edit, while non-EC editors can request edits on the talk page. As a contentious topic, Wikipedia administrators---a community-elected position---are empowered to place editing restrictions on disruptive accounts, without having to open a case at an administrative noticeboard. Participant 11 explained what happens when a topic is designated as contentious: 
\begin{quote}
    "\textit{The rules are sort of heightened. It's not necessarily that there any new rules, it's just that enforcement happens faster, and maybe more liberally, right? So we're cautioning editors to sort of like keep it within the lines more. Because you might get blocked or topic banned faster than you would in any other topic area.}"
\end{quote}

The editing restrictions in the topic area of the Russo-Ukrainian War increased over the course of the evolving conflict. The topic area of Eastern Europe has been designated as contentious since 2007, under the Wikipedia acronym WP:CT/EE.\footnote{\url{https://en.wikipedia.org/wiki/Wikipedia:Contentious\_topics/Balkans\_or\_Eastern\_Europe}} When the article "Russian invasion of Ukraine" was created on February 24, 2022, it was already eligible for protection as a contentious topic under Eastern Europe. Within 12 hours of the article's creation, a request for EC page protection was submitted and quickly approved. As the conflict progressed, other articles about the war were similarly protected: de facto contentious under the Eastern Europe topic area and EC protected on an ad-hoc basis. 

In October 2022, restrictions were increased further after Wikipedia editors enacted community authorized sanctions. Under General Sanctions/Russo-Ukrainian War, all war-related articles are EC protected, non-EC editors cannot create new articles, and they cannot participate in internal project discussions on talk pages or noticeboards. Known by the acronym WP:GS/RUSUKR, these community-enacted general sanctions are among some of the most strict protections on Wikipedia.\footnote{\url{https://en.wikipedia.org/wiki/Wikipedia:General_sanctions/Russo-Ukrainian_War}} Having all articles within a topic area be EC protected is only seen in four other areas on English-language Wikipedia: Israel-Palestine, Armenia-Azerbaijan, Kurds and Kurdistan, and antisemitism in Poland.

\textcolor{black}{The ability to limit who can edit articles has existed for several years, a result of Wikipedia editors dealing with contentious topics previously. Six levels of page restrictions exist: pending changes, semi, extended confirmed, template, full, and interface.\footnote{\url{https://en.wikipedia.org/wiki/Wikipedia:Protection_policy\#Comparison_table}} Extended confirmed (EC) level was the most heavily employed in the Russo-Ukrainian topic area. Several participants noted that EC page protections evolved from, in part, dealing with the contentious topic of Israel-Palestine.\footnote{\url{https://en.wikipedia.org/wiki/Wikipedia:Contentious\_topics/Arab-Israeli\_conflict}} For the Russo-Ukrainian War articles, the EC protections were widely described as successful by participants. Participant 09 said:}
\begin{quote}
   \textit{ “A lot of this stuff related to the war is so locked down with extended confirmed protection that I- I'm struggling to think of examples of [disruption]. I guess sometimes you'd maybe see stuff on the talk page of like IP users showing up and saying like, `This is biased. This is like so far off from truth. You people will never admit Ukraine lost' and that kind of thing.”}
\end{quote}

\textit{Moderating talk page discussions:} With page protections shielding articles, non-EC editors would often attempt to continue their disruption on the talk pages. Participants described various measures that were taken to keep talk page discussions constructive. At the least extreme, editors marked talk page discussions as closed and place a short explanation describing which Wikipedia policies or guidelines were violated. Going further, editors were able to collapse talk discussions so that its visibility was reduced, a practice know as `hatting' on Wikipedia.\footnote{\url{https://en.wikipedia.org/wiki/Wikipedia:Closing\_discussions\#HATTING}}

Under the community sanctions WP:GS/RUSUKR, non-EC editors are not allowed to engage in talk page discussions related to content disputes. One of Wikipedia's processes for dispute resolution is a Request for Comment (RfC), where editors can discuss content decisions.\footnote{\url{https://en.wikipedia.org/wiki/Wikipedia:Requests\_for\_comment}} On articles related to the Russo-Ukrainian war, non-EC editors attempted to disrupt these processes. Participant 05 described having to place a banner within the RfC discussion, notifying that comments from non-EC editors will be removed. Participant 05 recalled:
\begin{quote}
    "\textit{The banner hasn't always been present, but it's been acted upon. And then I've been putting banners in. Shortly after, we started becoming aware of how we could use this to manage all the, all the inexperienced and opinionated editors that were basically drive by.}"
\end{quote} 

Of the talk page moderation measures, one of the most extreme would be to fully EC-protect the talk page, not allowing non-EC editors to engage at all on the talk page. Participants noted that such an extreme measure is only enacted for short periods of time. To our knowledge, full EC-protections for talk pages have not been enacted on articles related to the Russo-Ukrainian War. Participant 11 explained, "Most of the time a talk page is not EC-protected, and individual editors and admins will just revert or delete participation by people who are not permitted because of the EC restriction." One participant noted they had only seen full EC talk page protections in the Israel-Palestine topic area. 

\textit{\textcolor{black}{Perennial reliable sources:}} \textcolor{black}{A central part of writing Wikipedia articles involves citing high-quality, reliable sources -- encapsulated in the core content policy Verifiability.\footnote{\url{https://en.wikipedia.org/wiki/Wikipedia:Verifiability}} Wikipedia maintains a list of perennial reliable sources, which summarizes the community's consensus on sources whose reliability has been repeatedly debated.\footnote{\url{https://en.wikipedia.org/wiki/Wikipedia:Reliable\_sources/Perennial\_sources}} Russian state media outlets such as RT, Sputnik, and TASS had been deemed ‘generally unreliable’ by the Wikipedia community prior to the February 2022 invasion. Select Russian state media, such as TASS, can be used only to attribute statements from the Russian government. Many participants referenced this virtual ban on Russian state media---as codified in the perennial sources list---as an effective measure that prevented Russian state-aligned narratives from appearing in war-related articles. Participant 07 commented on how Wikipedians ascertain the overall reliability of sources:} 
\begin{quote}
    "\textit{All sources are biased. Biased because of the people who own them. Biased because of the people who write for them. We have to accept that. Therefore we go by ‘how often this does this source tell outrageous lies?’ … this is one reason why we don't use Russian sources, because we know full well they tell outrageous lies.}"
\end{quote} 
\textcolor{black}{To illustrate this, Participant 07 noted how the Russian government had outlawed the use of terms "invasion" or "war” to describe the fighting in Ukraine, and as such Russian state media would be unfit for inclusion in a Wikipedia article~\cite{reuters_russia_2022}.}

\subsubsection{\textbf{Disruptive activity from new, unregistered users was typically simpler to address than disruptive activity from experienced users.}}
Participants recounted how both Russia-aligned and Ukraine-aligned accounts demonstrated varying levels of experience with Wikipedia: some accounts were new, unregistered accounts, while other accounts were more experienced and even had attained extended confirmed (EC) status. \textcolor{black}{While unregistered users were mostly impeded by Wikipedia's page protections, the more experienced, EC editors were able to change article content, and thus able to engage in protracted editing conflicts.}

Often referred to as IP users, as their edits are attributed to an IP address rather than a username, unregistered users misused the talk page to criticize the article and bad-mouth other editors. Participant 07 noticed IP users making the same point across multiple articles: "Across multiple pages there would be patterns where like, for example, the stuff about it being a special military operation to de-nazify Ukraine. That occurred over multiple pages, not just on the invasion of Ukraine page." Given that many Russo-Ukrainian War articles had been protected so that only more experienced editors could edit, the unregistered users were largely unsuccessful in changing article content. \textcolor{black}{Relegated to the talk pages, IP users would launch disruptive comments, but these were quickly archived or even deleted.} Participant 02 commented: "Often it would be people just coming and ranting, `I hate this article. It's rubbish. You're just awfully biased.' And it's not going to make any difference at all."

By contrast, experienced editors who engaged in disruptive activity were more difficult to counter. These editors exhibited an understanding of Wikipedia's policies and had reached extended confirmed (EC) status by having an account older than 30 days and having made over 500 edits. The experienced, EC editors were able to edit the protected Russo-Ukrainian War articles, rather than spam the talk pages as the unregistered users did. Disruptive EC editors created biased and low-quality content, using wiki-lawyering in their talk page arguments to keep their work from being reverted. Several participants noted how EC editors would create misleading citations in an attempt to support their biased content. Participant 08 recalled protracted disputes over citations used to demonstrate whether the Azov Brigade was neo-Nazi affiliated or not. Participant 08 noted: 
\begin{quote}
    "\textit{Now pro-tip, usually, if there's six references [for a single sentence], that means there has been a dispute, and it may be wrong.}" 
\end{quote}

\subsubsection{\textbf{\textcolor{black}{Wikipedia} editors consider the main invasion article to be of fairly high quality, but note that more peripheral articles can suffer from disruptive editing.}}
Participants tended to be in agreement that the main invasion article---Russian invasion of Ukraine---was fairly neutral and free of out-right bias given its high-visibility. Participants explained how an article viewed by many people also attracts a large number of editors---many of whom consider editing high-volume articles to be particularly impactful work on Wikipedia. Participant 06 observed, "It's a function of eyes. People have it on their watch list. And so it's very difficult to get away with casual fly-by vandalism on what is the main page." \textcolor{black}{Participant 02 commented:
\begin{quote}
    "\textit{I would say the attempts to insert misinformation, I think, are very hard to make work on a very highly trafficked article. Unless it's from someone who really understands Wikipedia very well. And by its nature, Wikipedia is quite a hard thing to learn.}"
\end{quote}}

Periphery articles related to the conflict, however, were often described as more vulnerable to disruptive and biased editing, given that fewer editors have these articles on their radar. Participant 10 explained, "We are talking about maybe dozens of articles altogether, which are on many watch lists. And we are talking about hundreds of 1000s of articles which are in the topic area, which are not on any watch list." Participant 10 added that while he watches some of the more peripheral articles related to Ukraine, he cannot watch them all, and that small disruptive edits---such as edit warring over the Russian or Ukrainian spelling of geographical places---will likely occur without anyone noticing. Participant 06 echoed: 
\begin{quote}
    "\textit{By the time you've gone down the 10\textsuperscript{th} level of a category tree ... you will find evidence of gaming, of partisan, of bias ... There are just literally 1000s of [articles] out there. It's very difficult to police them all.}" 
\end{quote}

Previous work has described this main versus periphery dichotomy. Hickman et al.~\cite{hickman_understanding_2021} interviewed Wikipedia editors who maintain articles related to the Kashmir region in the English-, Hindi-, and Urdu-language editions. Greenstein \& Zhu~\cite{greenstein_experts_2018} compared the quality of Wikipedia articles to Encyclopedia Britannica articles. Both papers found that the more attention an Wikipedia article received---i.e. number of active editors and revision count---the more neutral the article was. Both papers likened this phenomenon to the software development axiom Linus's law: "given enough eyeballs, all bugs are shallow." Our findings concur with this previous work that Linus's law can apply not only to software bugs, but also to content quality.\footnote{\url{https://en.wikipedia.org/wiki/Linus\%27s\_law}} 

\subsubsection{\textbf{\textcolor{black}{Wikipedia} editors are split in their perception of how well Wikipedia responded overall.}}
Participants differed in their assessments of how Wikipedia responded to the disruptive editing from Russia-aligned and Ukraine-aligned accounts. Six of the 13 editors we spoke with considered Wikipedia to overall have done a decent job in maintaining information integrity. These editors considered the extended-confirmed page protections to have stymied the bulk of disruptive editing. Participant 02 noted, "I wouldn't say it's cast iron protection. But I would say that out and out disinformation has a really hard time getting through that." Three editors were more lukewarm in their assessment, often noting that while page protections were helpful, the articles still faced other issues: disagreement over what sources were considered reliable and the time-sink of responding to disruptive behavior, i.e. enforcing bans and cleaning up low-quality articles.

Three editors had a mostly negative assessment of Wikipedia's response to disruptive editing in the Russo-Ukrainian topic area. These editors often described the main versus periphery article phenomenon, noting that while the main RIU article was in decent shape, many other periphery articles struggled with information integrity, such as articles about ongoing battles. These participants also considered the experienced EC editors who engaged in wiki-lawyering to be a considerable issue for the online encyclopedia. Participant 06 observed, "No, really, Wikipedia was not set up ... was not able to cope with the outfall from that war. It got too big, too quick, and it quickly emerged that their entire structure of Wikipedia guidelines, policies, was open to gamification." Participant 03 commented, "There's no requirement for Wikipedia editors to actually be neutral. There's just a neutral point of view policy, but that's for contents, not behavior ... I've become pretty jaded over Wikipedia lately." Finally, one participant demurred on their assessment, given that as an admin, they had approved requests for page protections, had not developed opinion on how the topic area was faring overall. 

\section{Discussion}

\subsection{Where did the trolls go?} Participants were hesitant to classify the disruptive editing they encountered as part of a state-backed information campaign from either Russia or Ukraine. Russia's efforts at waging information operations against Ukraine via social media, TV, and print media are well-documented~\cite{lucas_winning_2016}. Investigative journalism identified the Internet Research Agency, an internet troll farm located in Saint Petersburg, as the operators behind many Russian state information operations~\cite{carroll_st_2017, dossier_cyber_2023}. Following the 2022 invasion, scholarship has uncovered how Russian state media narratives have circulated on social media platforms, including Twitter, Facebook, Reddit, Telegram, and VKontakte~\cite{pierri_propaganda_2023, hanley_happenstance_2023, hanley_partial_2024, park_challenges_2022}. So why did we not find clear evidence of a Russia-aligned information operation on Wikipedia? There are several possibilities. 

One possibility is that the Russian government did not attempt any manipulation on English-language Wikipedia in the 2022-2023 time frame, having deprioritized the online encyclopedia in favor of other online targets. Scholarship has noted the extensive use of the encrypted messaging platform Telegram to promote Russia-aligned narratives around the war~\cite{stokel-walker_russias_2022, loucaides_kremlin_2023, aleksejeva_networks_2023, hanley_partial_2024}. For example, the Telegram channel "War on Fakes" was created on February 23, 2022---one day before the Russian invasion of Ukraine---and dismissed negative coverage of the Russian military as being faked, co-opting techniques from legitimate fact-checking sites. Investigative reporting has linked this Telegram channel to a journalist affiliated with Russian state media~\cite{bergengruen_unmasking_2023}. Beyond Telegram, reporting from BBC has exposed a Russian information operation on TikTok consisting of at least 800 accounts, which sought to discredit Ukrainian officials~\cite{robinson_ukraine_2023}. As mentioned in our findings, Participant 10 considered there to be paid Russia-aligned editors on Wikipedia in 2014-2016 but not in 2022-2023. As such, it is possible that Wikipedia is not presently a priority information target for the Russian government, as resources are being spent on trending social media platforms, such as Telegram and TikTok.

A second possibility is that the Russian government attempted to launch an information campaign on Wikipedia in the early days of the invasion, but later discontinued their efforts. As mentioned earlier, the RIU article was extended confirmed (EC) protected within 12 hours of its creation. Other Russo-Ukrainian War articles were also quickly EC-protected. As such, non-EC editors were left the ability to request edits on article talk pages. When these accounts did request an edit, they often provided links to Russian state media, which the Wikipedia community does not consider to be a reliable source. Moreover, participants emphasized that since the main war-related articles received such high levels of attention, disruption rarely went unnoticed. Wikipedia's barriers to entry for new editors, established reliable source policies, and heightened vigilance for war-related articles could have stymied an information campaign.
  
A third possibility is that the Russian government did orchestrate some part of the Russia-aligned disruptive editing described in our interviews, but purposefully designed the campaign to look sporadic and uncoordinated to avoid detection. Participant 01 observed that the Russia-aligned editors had varying writing styles and levels of English proficiency. Participant 06 described how in the contentious topic area of Armenia-Azerbaijan, the activity of disruptive editors evolved over time, from more crude to more subtle tactics as editors became more familiar with Wikipedia. It is possible that the Russian state government's tactics have evolved since the 2014-2016 paid editing activity described by Participant 10. A sign of this evolution might be the presence of experienced Russia-aligned editors who had achieved extended confirmed (EC) status. Participant 03 recalled encountering around 8-10 such accounts, while other participants mentioned around 2-3. Though the Russia-aligned EC editors appear to be small in number, participants concurred that responding to their disruption took considerable effort.

Relatedly, our participants could have been too generous in their "assume good faith" mindset---as the Wikipedia behavioral guideline encourages. While many participants considered the Russia-aligned editors to likely be nationalistic individuals, the English-language RIU article has been banned in Russia since May 2022.\footnote{\url{https://en.wikipedia.org/wiki/List\_of\_Wikipedia\_pages\_banned\_in\_Russia}} As such, when a talk page was spammed by Russia-aligned IP users, a good faith mindset assumes that these accounts were either Russian citizens violating their government's Wikipedia ban or individuals abroad sympathetic to Russia, rather than state-backed accounts given clearance to access a banned website. 

\subsection{\textcolor{black}{Connections to prior work on information integrity and Wikipedia}} \textcolor{black}{
A central finding of our paper is: while the Russo-Ukrainian War articles faced disruption, Wikipedia had policies and processes in place to respond and mitigate it. This finding aligns with and extends existing literature on Wikipedia and information integrity. McDowell \& Vetter \cite{mcdowell_it_2020} describe how Wikipedia's policies enable the encyclopedia to assuage problematic content:} 

\begin{quote}
\textcolor{black}{“Wikipedia’s battle against fake news, misinformation, and disinformation is waged within and through community-mediated practices, and policies put into place in the encyclopedia to verify and validate information, to ensure accuracy, neutrality, and to guard against bias and misinformation.”}
\end{quote}

\textcolor{black}{In our findings, page protections were central to our participants response to disruption. Editors have to be in good standing to edit, especially on sensitive topics like the Russia-Ukraine war. Hill \& Shaw \cite{hill_page_2015} describe page protections as a type of "hidden wikiwork" --- an unobtrusive feature which enables the encyclopedia to resolve disputes. Ajmani et al. \cite{ajmani_peer_2023} conceptualize page protections as a "frictions-mechanism" to safeguard information quality on the encyclopedia. McDowell \& Vetter \cite{mcdowell_it_2020} similarly note the importance of user access levels to deter vandalism and low quality editing. Both Wikipedians and researchers caveat, however, that page protections should be used sparingly and for limited amounts of time~\cite{ajmani_peer_2023, ruprechter_protection_2023}.\footnote{\url{https://en.wikipedia.org/wiki/Wikipedia:Protection_policy}} Over restricting access to a page can put strains on the amount of editors available to maintain those pages, potentially leading to reductions in information quality.}

\textcolor{black}{Reliable sources were also frequently mentioned by our participants. McDowell \& Vetter \cite{mcdowell_it_2020} discuss the importance of reliable sources in maintaining Wikipedia’s information integrity. They aptly note that while the policy of Verifiability is fairly short --- “articles must be based on reliable, independent, published sources with a reputation for fact-checking and accuracy” \cite{mcdowell_it_2020} --- this policy page is supplemented by dozens of supporting pages, which provide guidelines on how to arbiter whether a source is reliable. One of these supporting pages is the Perennial Reliable Sources list --- which our participants described as important to responding to disruption on the Russia-Ukraine War articles. Steinsson \cite{steinsson_rule_2023} discusses how this "sourcing hierarchy" was developed overtime, as Wikipedia editors sought to delineate fringe views from mainstream ones. In addition, prior work has shown the presence of numerous and high-quality sources is important to how Wikipedia readers assess an article’s credibility \cite{elmimouni_why_2022}}.

\textcolor{black}{Beyond page protections and reliable sources, our participants mentioned other Wikipedia features which have been studied previously. Participants noted the use of automated Wikipedia tools to respond to disruption on the Russo-Ukrainian War articles, such as reporting of potential vandals, which has been discussed in previous CSCW work \cite{geiger_work_2010}. Participants also described differing information integrity outcomes, with main articles receiving more attention and thus being of higher-quality than peripheral articles. This main versus periphery dichotomy has been described in at least two previous studies~\cite{greenstein_experts_2018, hickman_understanding_2021}. Finally, participants also frequently cited the importance of the neutral point of view policy (NPOV) in guiding content decisions on contested articles. Steinsson \cite{steinsson_rule_2023} argues that the NPOV policy has enabled Wikipedia to become an increasingly reliable information source since its creation in 2001.}

\subsection{Wikipedia is not social media. Perhaps social media could be more like Wikipedia?} Across the interviews, participants considered Wikipedia to be more successful than social media at preventing the spread of false or misleading information during evolving geopolitical events, including the Russo-Ukrainian War. Several participants attributed this success to Wikipedia's higher barriers to entry, \textcolor{black}{noting how learning to edit Wikipedia required time and patience.} In addition to high barriers to entry, Participant 12 emphasized the role of the Wikipedia community in resolving content and conduct disputes: "Our processes come from the community and they're operated by the community. At Twitter, Facebook, or wherever, you know, that's all done largely top down ... And they don't, they're not really under any obligation to explain their decisions all that well." As our participants acknowledged, Wikipedia is an attractive target for information operations, but unlike social media, Wikipedia appears to be adept at preventing false or misleading information from entering its articles.

\textcolor{black}{Among social media platforms, content moderation strategies range from centralized to decentralized. Wikipedia is a combination of these approaches, where moderation is crowd-sourced but based on a universal set of rules. For social media platforms that employ a centralized approach, Trust and Safety teams review content for violating company-defined rules (e.g. X, formerly Twitter). This approach can run into issues of being overly opaque and devoid of community input. For social media platforms that employ a decentralized approach, content is moderated by the community, often with community-defined rules (e.g. Reddit, Discord). This approach can lead to differing moderation decisions across one platform, in addition to smaller sub-communities being overwhelmed by the demands of moderation. A recent interview study found that Reddit moderators wanted more platform-level guidance when addressing ambiguous cases of COVID-19 misinformation~\cite{bozarth_wisdom_2023}. Wikipedia offers a middle-ground approach, where there is a universal set of rules that are enforced in a crowd-sourced manner by community members. While there is no one-size-fits-all approach to content moderation, social media platforms could look to Wikipedia for how to increase both the transparency and the community engagement of its moderation processes.}

While our interview protocol did not explicitly include design-oriented questions, our qualitative analysis suggests several design implications from Wikipedia that might be of use to social media platforms. Wikipedia benefits from experienced `old-timer' editors who pass on social norms to incoming users, such as civility and assume good faith~\cite{forte_scaling_2008}. One idea would be for a social media platform to invite long-term users to help create content and conduct policies for the community. Or, to elevate them to these roles via some mechanism. Social media sites Reddit and Discord already employ community-based content moderation. In addition to community creation of policies, community enforcement might also be beneficial for social media sites, in the spirit of Linus' Law---with enough eyes, most problematic content should be flagged. To thank social media users for their work in helping maintain online spaces, social media platforms could learn from Wikipedia's Barnstars: tokens of appreciation that editors can send to other editors for their contributions.\footnote{\url{https://en.wikipedia.org/wiki/Wikipedia:Barnstars}} In addition, social platforms could look to the success that Wikipedia has had in debating and curating lists of reliable and unreliable sources, a process which could be replicated in other social platforms. At minimum, platforms could import the work that Wikipedians have done, and label or algorithmically alter content from low-reliability sources.

\subsection{\textcolor{black}{Limitations of studying only English-language Wikipedia}} \textcolor{black}{The focus of our study on the English-language edition of Wikipedia, one of over 340 language editions,\footnote{\url{https://meta.wikimedia.org/wiki/List_of_Wikipedias}} presents several limitations. First, the English-language edition of Wikipedia is arguably one of the most well-resourced editions, which might make it singularly resilient to disruptive editing. Previous work has proposed that active editors, community governance, and diversity in demographics all contribute to making Wikipedia language editions resilient to knowledge integrity risks~\cite{aragon_preliminary_2021, saez-trumper_online_2019}. Kharazian et al. ~\cite{kharazian_governance_2023} investigated how differing community governance structures can lead to different outcomes: the Croatian language edition of Wikipedia was captured by far-right editors motivated by a nationalistic agenda, while the Serbian and Serbo-Croation language editions remained resilient to governance capture.}  

\textcolor{black}{Second, while Wikipedia's policies and processes are similar across language editions, they are not identical~\cite{johnson_considerations_2022}. Page protections --- an aspect of Wikipedia central to maintaining English-languages articles related to the Russo-Ukrainian --- exist in other language editions with mild variations. For example, the German-language edition has four types of page protections, as opposed to six types in the English-language edition, and users report misconduct of other users, rather than request a page protection outright~\cite{ruprechter_protection_2023}. Talk pages --- another aspect of Wikipedia discussed extensively by our participants --- is a forum where  edits are first discussed before the article is updated, or at least for English-language Wikipedia. Bipat et al. \cite{bipat_we_2018} found that the Spanish-language and French-language editions do not use the talk page to discuss potential article edits. Given these varying policies and processes, there is potential for heterogeneity in the responses to Russia-Ukraine War articles across the different language editions.} 

\textcolor{black}{Third, the Russian-language edition of Wikipedia has faced attempts from the Russian government to censor its articles --- warranting further study. On March 1, 2022, the Russian agency for the Supervision of Communications, Information Technology and Mass Media (Roskomnadzor) asked the Wikimedia Foundation to remove the Russian-language article "Russian invasion of Ukraine."~\cite{gregory_russia_2022} In the proceeding months of the war, Roskomnadzor continued to demand Wikipedia articles be removed with threats of fines and began placing articles on a list of forbidden sites.\footnote{\url{https://en.wikipedia.org/wiki/List_of_Wikipedia_pages_banned_in_Russia}} In June 2023, Ruwiki was launched, a government-sanctioned fork of Russian Wikipedia, which copied over 1.9 existing articles while removing any mention of the invasion of Ukraine~\cite{corfield_russia_2023}. Such attempts at censorship could explain why our participants did not perceive there to be any state-backed information campaign: perhaps the Russian government was more focused on censoring Wikipedia than influencing it.}

\textcolor{black}{Despite these limitations, it is important to note the utility of studying English-language Wikipedia for non-English language contexts. Our participants noted how editors from all over the world, many of whom English is not their first language, focus their time and energy on the English-language edition given its high visibility and reach. For several of our participants, this was indeed the case: English was not their first language, yet they contributed extensively to English-language articles on the Russo-Ukrainian War. As such, we emphasize that our study's findings are not restricted to only native English-language editors or readers on Wikipedia.}

\subsection{Future work}
We believe this work could be informative for future, large-scale observational studies of Wikipedia. Across interviews, participants frequently mentioned other topic areas, such as Israel-Palestine and Armenia-Azerbaijan, and suggested potential for a comparative study of the handling of contentious topics on Wikipedia. \textcolor{black}{A comparative study of contentious topics might further probe the extent of page protections' effectiveness: are there perhaps instances when an information operation has been successfully waged even with these protections in place?} \textcolor{black}{Future work could also investigate how the pressures of an authoritarian regime impacts the governance of a Wikipedia edition. Relatedly, a comparison of the Russian-language edition of Wikipedia and its government-sanctioned fork Ruwiki could be revealing: what content has been wiped from Ruwiki; do Ruwiki editors collaborate like editors in the original Russian-language edition?} Another approach would be to explore the coverage of the Russo-Ukrainian war in other language Wikipedias. As our participants noted, if Wikipedia is indeed a reflection of what the mainstream sources say, then perhaps Hindi Wikipedia\footnote{\url{https://hi.wikipedia.org/wiki/}} or Italian Wikipedia,\footnote{\url{https://it.wikipedia.org/wiki/}} countries with closer diplomatic relationships with Russia, might present the Russo-Ukrainian War differently than English Wikipedia.

\section{Conclusion}
Across 13 interviews with expert Wikipedia editors, we surfaced challenges faced by articles related to the Russo-Ukrainian War \textcolor{black}{on the English-language edition}. Our participants did not perceive there to be clear evidence of a state-backed information campaign, a finding that stands in contrast to scholarship showing evidence of Russia-aligned information operations targeting social media platforms. Whether or not a state actor was present, participants reported high-levels of disruptive editing from both Russia-aligned and Ukraine-aligned accounts, which created time-intensive maintenance work for editors. \textcolor{black}{The English-language edition of} Wikipedia overall appeared prepared to address the disruption in the wake of the Russia-Ukraine War, relying upon existing policies and processes honed in other contentious topic areas. We are optimistic this paper will support future work to further elucidate Wikipedia’s resilience in the face of information manipulation online and explore potential lessons for other internet platforms.

\bibliographystyle{ACM-Reference-Format}
\bibliography{references}

\appendix
\section{Interview Protocol}
\subsection{Involvement with Wikipedia}
\begin{enumerate}
    \item How did you start editing for Wikipedia?
    \item How did you become involved in the editing articles related to the Russian invasion of Ukraine?
    \begin{enumerate}
        \item Is this your main topic area, or are you more involved in other topic areas on Wikipedia?
    \end{enumerate}
\end{enumerate}
\subsection{Experiences editing in the Russo-Ukrainian War topic area}
\begin{enumerate}
    \item Walk me through what it has been like working on the articles related to the war.
    \item Can you think of a case where there was problematic or disruptive behavior on an article?
    \begin{enumerate}
        \item What was your process or strategy to determine whether the behavior was disruptive?
    \end{enumerate}
    \item Can you think of a case where there was problematic or disruptive behavior on a talk page?
    \begin{enumerate}
        \item What was your process or strategy to determine whether the behavior was disruptive?
    \end{enumerate}
    \item How do you decide whether you'll respond to the disruption?
    \begin{enumerate}
        \item If so, what does a response look like?
        \item What types of tools do you have at your disposal in order to respond?
        \item How much does the intent of the account matter in terms of the what action is taken?
    \end{enumerate}
    \item What do you make of the disruptive activity you have seen? Do you consider it to be small in scope, or have you come across things that look more like larger, coordinated attacks?
    \begin{enumerate}
        \item How do you make an assessment as to whether the disruption was coordinated?
        \item If coordinated, how would you describe the goal(s)/aim(s) of the coordination?
    \end{enumerate}
\end{enumerate}
\subsection{Wikipedia's response to disruption in the Russo-Ukrainian War topic area}
\begin{enumerate}
    \item How do you think Wikipedia’s processes, policies, and norms have held up for maintaining articles about an on-going war?
    \begin{enumerate}
        \item Can you think of a case that worked well and a case that did not work well?
        \item For the cases that didn't work, in what ways did they fail?
        \item To my understanding, only extend confirmed editors can edit articles in this topic area, due to WP:GS/RUSUKR. Do you think the general sanctions were effective or not effective for responding to disruptive activity?
    \end{enumerate}
    \item In the archived talk pages, I noticed that many discussions centered around Wikipedia policies. Can you think of a time when a policy was very relevant to resolving a dispute?
    \begin{enumerate}
        \item Which policies were most effective for responding to disruptive activity?
        \item Were any policies consistently debated? Were any policies consistently violated?
        \item What do you consider to be a reliable source? How do you make this judgement?
    \end{enumerate}
    \item How does your experience with the Russian invasion of Ukraine compare with other topics areas on Wikipedia?
\end{enumerate}
\end{document}